\documentclass[conference]{IEEEtran}
\usepackage{amsmath,amsfonts}
\usepackage{graphicx,amssymb}
\usepackage{algorithmic,algorithm}
\usepackage{amsthm,dsfont}
\usepackage{subfigure,url}
\usepackage[noadjust]{cite}
\usepackage{xcolor}
\usepackage{colortbl,booktabs}

\theoremstyle{plain}

\newtheorem{thm}{Theorem}

\theoremstyle{remark}

\newcounter{longequ}[longequ]
\addtocounter{longequ}{11}

\allowdisplaybreaks

\IEEEoverridecommandlockouts
\begin{document}
%
\title{A Low-Complexity Algorithmic Framework for Large-Scale IRS-Assisted Wireless Systems}
\author{\IEEEauthorblockN{Yifan Ma\IEEEauthorrefmark{1},
		Yifei Shen\IEEEauthorrefmark{1},  Xianghao Yu\IEEEauthorrefmark{2}, Jun Zhang\IEEEauthorrefmark{3}, S.H. Song\IEEEauthorrefmark{1}\IEEEauthorrefmark{4}, and Khaled B. Letaief\IEEEauthorrefmark{1}}\\
	\IEEEauthorblockA{\IEEEauthorrefmark{1}Dept. of ECE, The Hong Kong University of Science and Technology, Hong Kong\\
	\IEEEauthorrefmark{2}Friedrich-Alexander-Universit\"{a}t Erlangen-N\"{u}rnberg, Germany\\
	\IEEEauthorrefmark{3}Dept. of EIE, The Hong Kong Polytechnic University, Hong Kong\\
	\IEEEauthorrefmark{4}Division of ISD, The Hong Kong University of Science and Technology, Hong Kong\\
	Email: \IEEEauthorrefmark{1}\{ymabj, yshenaw, eeshsong, eekhaled\}@ust.hk,  \IEEEauthorrefmark{2}xianghao.yu@fau.de, \IEEEauthorrefmark{3}jun-eie.zhang@polyu.edu.hk}}
\maketitle

\begin{abstract}
Intelligent reflecting surfaces (IRSs) are revolutionary enablers for next-generation wireless communication networks, with the ability to customize the radio propagation environment. To fully exploit the potential of IRS-assisted wireless systems, reflective elements have to be jointly optimized with conventional communication techniques. However, the resulting optimization problems pose significant algorithmic challenges, mainly due to the large-scale non-convex constraints induced by the passive hardware implementations. In this paper, we propose a low-complexity algorithmic framework incorporating alternating optimization and gradient-based methods for large-scale IRS-assisted wireless systems. The proposed algorithm provably converges to a stationary point of the optimization problem. Extensive simulation results demonstrate that the proposed framework provides significant speedups compared with existing algorithms, while achieving a comparable or better performance.
\end{abstract}

\IEEEpeerreviewmaketitle

\section{Introduction}

Recently, intelligent reflecting surfaces (IRSs) have emerged as an energy-efficient and powerful approach for reconfiguring the wireless propagation environment through programmable reflection \cite{wu2019towards}. Equipped with a large number of low-cost \emph{passive} reflective elements, e.g., phase shifters and dipoles, IRSs are able to enhance the quality of the received signals with limited power consumption and radio frequency (RF) chains \cite{huang2019reconfigurable}. Furthermore, IRSs can be embedded into existing environmental objects, e.g., the facades of buildings and smart t-shirts \cite{di2019idea}, which greatly reduces the implementation cost. To sum up, IRSs are key enablers of next-generation wireless communications by relieving the burdens on hardware expenditure and energy cost, given their remarkable ability to customize the radio propagation environment.

The highly non-convex unit modulus constraints induced by the implementation of phase shifters, however, pose great algorithmic challenges. Early attempts resorted to semidefinite relaxation (SDR)-based methods \cite{cui2019secure}, which yield an approximate solution without optimality guarantee. To overcome this drawback, optimization methods on the complex circle manifold were developed in \cite{yu2019miso} and the first-order optimality is guaranteed. Nevertheless, a common limitation of SDR and manifold optimization is the high computational cost. Specifically, in SDR, solving a series of high-dimensional semidefinite programming (SDP) problems increases the computational burden drastically. In the manifold optimization-based algorithms, the nested loop architecture slows down the convergence \cite{yu2016alternating}. Therefore, existing methods for optimizing IRS-assisted systems are only applicable to small-scale IRSs. However, it is shown in \cite{DFRelay} that hundreds of reconfigurable elements are required to be competitive with conventional decode-and-forward relaying, when minimizing the total transmit power or maximizing the energy efficiency. It has motivated the development of low-complexity algorithms for large-scale IRSs in recent studies, e.g., fixed point method \cite{yu2019miso}, which, however, are problem-dependent. These problem-specific approaches require a laborious process and much prior knowledge of the considered optimization problems. In other words, how to develop a general design methodology for large-scale IRSs is still an open problem.

In this paper, we propose a low-complexity algorithmic framework for large-scale IRS-assisted systems. Specifically, we first employ alternating optimization (AO) to decouple the optimization variables and then leverage a first-order method for phase optimization. In contrast to the SDR and manifold optimization-based methods, this framework avoids convex relaxation, does not need to solve high-dimensional SDP problems, and gets rid of the nested loops. Thus, it significantly improves the computational efficiency. Different from the classic gradient descent (GD) method, the step size is delicately designed by taking the overall AO procedures into account to achieve a faster convergence rate. We then prove that a subsequence of the proposed algorithm is guaranteed to converge to a stationary point, even when the objective does not have a Lipschitz continuous gradient. We take IRS-enabled secure wireless communications and weighted sum-rate maximization in IRS-empowered systems as two applications. Simulations demonstrate that the proposed framework achieves a significant speedup compared with the SDR-based method, the manifold optimization-based method, and problem-specific low-complexity block coordinate descent (BCD) algorithms, while maintaining a comparable or better performance. 

\begin{table*}[t]	
\centering
\newcommand{\tabincell}[2]{\begin{tabular}{@{}#1@{}}#2\end{tabular}}
\caption{Typical Examples of the General Formulation \eqref{general}.} 
\resizebox{1\textwidth}{!}{
\begin{tabular}{|c|c|c|c|}
\hline
Problem & Objective function ${f}$ & Conventional communication techniques $\bf{Q}$ & Set constraint ${\cal X}_1$ \\ \hline
Spectral efficiency maximization \cite{yu2019miso} & \tabincell{c}{
Achievable spectral efficiency in \\ single-user MISO systems
} & ${\bf{w}}$: Transmit beamforming & ${\left\| {\bf{w}} \right\|^2} \le P $  \\ \hline
Secrecy rate maximization \cite{Yu2020Robust} & Sum secrecy rate of $K$ legitimate users & 
\tabincell{c}{
${\bf{w}}_k$: Transmit beamforming, \\ ${\bf{Z}}$: Artificial noise
}
& \tabincell{c}{
$\sum_{k=1}^{K}\left\|\mathbf{w}_{k}\right\|^{2} + {\rm{Tr}({\bf{Z}})} \leq P$,  \\ ${\bf{Z}} \succeq {\bf{0}}$
}   \\ \hline
Weighted sum-rate maximization \cite{guo2020weighted} & Weighted sum-rate of $K$ mobile users & ${\bf{w}}_k$: Transmit beamforming & $\sum_{k=1}^{K}\left\|\mathbf{w}_{k}\right\|^{2} \leq P $ \\
\hline
\end{tabular}
}
\label{example}
\end{table*}

\emph{Notations:} $x$ is scalar, ${\bf{x}}$ is vector, and ${\bf{X}}$ is matrix. 
Let ${{\bf{X}}^T}$, ${\bf{X}^*}$, and ${{\bf{X}}^H}$ denote the transpose, conjugate, and conjugate transpose of matrix X, respectively. 
${\mathop{\rm diag}\nolimits} \left( {{x_1}, \cdots ,{x_n}} \right)$ represents a diagonal matrix with entries ${{x_1}, \cdots ,{x_n}}$ on its main diagonal. 
${{\bf{I}}_M}$ stands for $M \times M$ identity matrix. 
Operation ${\mathop{\rm Re}\nolimits} \{ {\bf{X}}\}$ constructs a matrix by extracting the real parts of the entries of matrix ${\bf{X}}$ while operation $\angle ({\bf{X}})$ extracts the phases of elements of ${\bf{X}}$. 
The modulus of a complex number is denoted by $\left|  \cdot  \right|$ and $j = \sqrt { - 1}$ is the imaginary unit. 
$\mathbb{C}^{m \times n}$ represents the set of all ${m \times n}$ complex-valued matrices. 
The Hadamard product is denoted by $\odot$.

\section{General Formulation and Low-Complexity Algorithmic Framework}
\par In this section, we first provide a general formulation for the optimization of IRS-empowered wireless communications and then present a low-complexity algorithmic framework incorporating alternating optimization and gradient descent.
\subsection{General Formulation}
We consider the following general formulation of IRS-assisted systems

\begin{equation} \label{general}
\begin{aligned}
\min_{{\bf{Q}},{\bf{\Phi }}} \quad & {f}({\bf{Q}},{\bf{\Phi }}) \\
\mbox{s.t.}\quad
&\left| {{\bf{\Phi }}_{i,i}} \right| = 1, \quad \forall i, \\
&{\bf{Q}} \in {{\cal X}_1},
\end{aligned}
\end{equation}
which typically involves two blocks of variables. $\bf{Q}$ includes variables of conventional communication techniques, e.g., the beamforming vector and artificial noise, while ${\bf{\Phi}} = {\mathop{\rm diag}\nolimits} (e^{j\theta_1},\cdots,e^{j\theta_M})$ is the phase shift matrix where $\theta_i$ denotes the phase shift of the $i$-th reflective element of the IRS. ${f}({\bf{Q}},{\bf{\Phi }})$ denotes the performance metric to optimize, e.g., the achievable rate or energy efficiency, ${\cal X}_1$ denotes the additional constraints for $\bf{Q}$, such as the transmit power constraint, and $\left| {{\bf{\Phi }}_{i,i}} \right| = 1$ represents the unit modulus constraint of the IRS. Table \ref{example} lists several example applications of the general formulation in the literature. To the best of the authors' knowledge, there is no general approach to obtain the optimal solution to such problems, mainly because of the deeply coupled variables and non-convex constraint. Hence, a practical algorithm design is of great importance.


\par To tackle the coupling of the optimization variables, alternating optimization is often adopted, by iteratively optimizing one set of variables while fixing the other variables \cite{cui2019secure,Alex2019secure}. In particular, in each iteration, we first update ${\bf{Q}}$ with ${\bf{\Phi }}$ fixed, and then optimize ${\bf{\Phi }}$ given a fixed ${\bf{Q}}$. These two steps are defined as two block updates in one iteration. When the phase shift matrix is given, the original problem reduces to a conventional communication problem without IRS, which has been investigated for decades and for which compelling mechanisms exist. The bottleneck of the whole optimization problem thus lies in the phase shift update part.

\par In the $t$-th iteration and for fixed ${\bf{Q}} = {\bf{Q}}^{(t)}$, the optimization problem becomes
\begin{equation} \label{eq2}
\begin{aligned}
\min_{{\bf{\Phi }}} \quad & {f}({\bf{\Phi }} | {\bf{Q}}^{(t)}) \\
\mbox{s.t.}\quad
&\left| {{\bf{\Phi }}_{i,i}} \right| = 1, \quad \forall i.
\end{aligned}
\end{equation}
For simplicity, ${\bf{\Phi }}$ can be rewritten as
\[{\bf{\Phi }} = {\mathop{\rm diag}\nolimits} \left({e^{j{\theta _1}}},{e^{j\theta 2}},\cdots,{e^{j{\theta _M}}} \right) = {U}({\bf{\Theta}}),\]
where ${\bf{\Theta}} = {[{\theta _1},{\theta _2},\cdots,{\theta _M}]^T} = \angle ({\bf{\Phi }})$, ${U}( \cdot )$ is a mapping such that ${U}({\bf{\Theta}})={\bf{\Phi}}$, and $M$ denotes the total number of non-zero entries in ${\bf{\Phi }}$. As a result, Problem \eqref{eq2} can be recast into
\begin{alignat}{2}
\min_{{\bf{\Theta}}} \quad & {f}({U}({\bf{\Theta}}) | {\bf{Q}}^{(t)}),  &{}& \label{eq3}
\end{alignat}
which becomes an unconstrained optimization problem and the effective gradient descent method is favorable for \eqref{eq3}.

\subsection{Gradient Descent Method}
\par Recent studies demonstrated that the first-order methods, e.g., gradient descent, are computationally-efficient and can often find \emph{globally} optimal solutions for problems with unit modulus constraints \cite{NIPS2018dual, shen2020complete}. This inspires us to apply variants of GD to solve \eqref{eq3}. The update rule of ${\bf{\Theta }}$ using GD is given by
\begin{equation}
{{\bf{\Theta }}^{(t + 1)}} = {{\bf{\Theta}}^{(t)}} - \gamma^{(t)} {\nabla _{\bf{\Theta}}}f({\bf{X}}^{(t)}),
\end{equation}
where ${\bf{X}}^{(t)}=({\bf{Q}}^{(t)}, {\bf{\Phi}}^{(t)})$, ${\nabla _{\bf{\Theta}}}f({\bf{X}}^{(t)})$ denotes the gradient 
in the $t$-th iteration, and the positive scalar $\gamma^{(t)}$ denotes the step size.
With the negative gradient being the descent direction, the selection of an appropriate step size is essential to promote convergence from remote initial points. Conventional steepest descent tries to identify the global minimizer of the objective function along the descent direction, which is too computational-expensive. A more practical strategy is performing Armijo-Goldstein (AG) line search \cite{numerical2006} to make $\gamma^{(t)}$ satisfy
\begin{equation} \label{eq:ag}
\begin{aligned}
&f({U}({{\bf{\Theta }}^{(t)}} - {\gamma^{(t)}}{\nabla _{\bf{\Theta }}}f({{\bf{X}}^{(t)}})),{\bf{Q}}^{(t)})  \\
& \le f({U}({{\bf{\Theta }}^{(t)}}),{\bf{Q}}^{(t)}) - c{\gamma^{(t)}}{\nabla _{\bf{\Theta }}}f{({{\bf{X}}^{(t)}})^T}{\nabla _{\bf{\Theta }}}f({{\bf{X}}^{(t)}}),
\end{aligned}
\end{equation}
where $0 < c < 1$ is a constant.
AG line search enforces a sufficient decrease in each block update of the phase shifts in \eqref{eq3}, which, however, may be too restrictive for solving the joint optimization problem \eqref{general}. 
To overcome this drawback, we develop a tailored step size chosen scheme that takes advantage of the joint alternating optimization procedures in tackling problem \eqref{general} while enjoying the convergence guarantee.

\par Recall that we first optimize $\mathop{{\rm{min}}}\limits_{{\bf{Q}} \in {{\cal X}_1}} {f}({{\bf{Q}} | {\bf{\Phi}}^{(t)}})$ with given ${\bf{\Phi}}^{(t)}$ and traditional methods could identify a high-performant solution to this subproblem.
This property benefits us to adopt a more aggressive step size in \eqref{eq3} and ensure an \emph{iteration-wise} monotonic decrease of the objective value rather than \emph{block-wise} monotonicity in \eqref{eq:ag}. Specifically, step size $\gamma^{(t)}$ in our framework is designed such that
\begin{equation} \label{eq4}
\begin{aligned}
& f({U}({{\bf{\Theta }}^{(t)}} - {\gamma^{(t)}}{\nabla _{\bf{\Theta }}}f({{\bf{X}}^{(t)}})),{\bf{Q}}^{(t)}) \\
& \le \underbrace{f({U}({{\bf{\Theta }}^{(t)}}),{\bf{Q}}^{(t-1)})}_{(a)} - c{\gamma^{(t)}}{\nabla _{\bf{\Theta }}}f{({{\bf{X}}^{(t)}})^T}{\nabla _{\bf{\Theta }}}f({{\bf{X}}^{(t)}}).
\end{aligned}
\end{equation}
The left-hand side in \eqref{eq4} denotes the objective value in the $t$-th iteration and (\ref{eq4})(a) is that in the $(t-1)$-th iteration. 
The second term in the right-hand side of \eqref{eq4} enforces the reduction of $f$ to be proportional to both the step size ${\gamma^{(t)}}$ and the gradient, which helps guarantee the convergence property. 
In practice, we progressively approach the largest feasible ${\gamma ^{(t)}}$. Namely, starting from $\gamma_0>0$, if the condition in \eqref{eq4} is violated, the step size is decreased by a factor of $0<\beta<1$. 

\par To conclude, unlike conventional methods that restrict a certain extent decrement in each block, our computationally-efficient approach guarantees the decrement on an iteration basis, which leads to a larger step size. The fast convergence speed will be shown empirically in Section \uppercase\expandafter{\romannumeral4}. The overall algorithm is summarized in \textbf{Algorithm 1} and its convergence properties are given in the following theorem.

\begin{algorithm}[htbp]
\caption{Alternating Optimization Based on Gradient Descent (AO-GD)}
\label{alg1}
\begin{algorithmic}[1]
\STATE {Construct an initial ${\bf{\Phi }}^{(0)}$ and let ${{\bf{\Theta }}^{(0)}} = \angle ({\bf{\Phi }}^{(0)})$. Set an initial step size $\gamma_0$ and a step size decay factor $\beta$. Let $t=0$;} 
\REPEAT 
\STATE {Fix ${\bf{\Phi }}^{(t)}$ and optimize ${\bf{Q}}^{(t)}$;} \label{a3}
\STATE {Fix ${\bf{Q}}^{(t)}$ and compute the gradient ${\nabla _{\bf{\Theta}}}f({\bf{Q}}^{(t)},{\bf{\Phi }}^{(t)})$;} \label{a4}
\STATE {Set step size $\gamma \leftarrow \gamma_0$;}
\WHILE{$t \ne 0$ and $f({U}({{\bf{\Theta }}^{(t)}} - \gamma {\nabla _{\bf{\Theta }}}f({{\bf{X}}^{(t)}})),{\bf{Q}}^{(t)}) > f({\bf{\Phi }}^{(t)}, {\bf{Q}}^{(t - 1)})- c\gamma {\nabla _{\bf{\Theta }}}f{({{\bf{X}}^{(t)}})^T}{\nabla _{\bf{\Theta }}}f({{\bf{X}}^{(t)}})$} 
\STATE {$\gamma \leftarrow \beta  \times \gamma $;}
\ENDWHILE
\STATE {${{\bf{\Theta }}^{(t + 1)}} \leftarrow {{\bf{\Theta}}^{(t)}} - \gamma {\nabla _{\bf{\Theta}}}f({\bf{X}}^{(t)})$;}
\STATE {${\bf{\Phi}}^{(t + 1)} \leftarrow {U}({{\bf{\Theta}}^{(t + 1)}})$;}
\STATE {$t \leftarrow t + 1$;}
\UNTIL{convergence.} 
\end{algorithmic}
\end{algorithm}

\begin{thm}\label{thm1}
Every limit point of the sequence $\{ {{\bf{X}}^{(t)}}\}$ generated by \textbf{Algorithm 1} is a stationary point of Problem \eqref{general}.
\end{thm}
\begin{IEEEproof}
Please refer to the Appendix. The argument is built on the proof of \cite[Theorem 1]{InexactBCD}.
\end{IEEEproof}


The key procedures in \textbf{Algorithm 1} are Steps 3 and 4. Step 3 can be solved through conventional methods, and Step 4 involves computing the gradient of the objective function. To show the generality of the proposed framework, we demonstrate two applications in the following section and show how to implement these two key steps for different application scenarios.

\section{Applications of The Proposed Framework}
In this section, two different problems for IRS-aided wireless communication systems are studied, namely, secrecy rate maximization \cite{Alex2019secure} and weighted sum-rate maximization \cite{guo2020weighted}.
\subsection{Secure Wireless Communications via IRS}
\par Consider an IRS-assisted secure wireless communication system where one transmitter equipped with $N_t$ antennas serves one single-antenna legitimate user, with the existence of a single-antenna eavesdropper and an IRS consisting of $M$ elements. The objective is to maximize the secrecy rate of the system by jointly optimizing the transmit beamformers and phase coefficients of IRS elements. The problem formulation is as follows:
\begin{equation} \label{secure}
\begin{aligned}
\max_{{\bf{w}},{\bf{\Phi }}} \quad & {f}({\bf{w}},{\bf{\Phi }}) = \frac{{1 + \frac{1}{{\sigma _{\rm{l}}^2}}{{\left| {{\bf{h}}_{\rm{l}}^H{\bf{\Phi Gw}}} \right|}^2}}}{{1 + \frac{1}{{\sigma _{\rm{e}}^2}}{{\left| {{\bf{h}}_{\rm{e}}^H{\bf{\Phi Gw}}} \right|}^2}}} \\
\mbox{s.t.}\quad
&{\bf{\Phi }} = {\mathop{\rm diag}\nolimits} \left( {{e^{j{\theta _1}}},{e^{j{\theta _2}}}, \cdots ,{e^{j{\theta _M}}}} \right),\\
&{\bf{w}} \in {{\cal X}_1} = \left\{ \left. {{\bf{w}}_0} \right| {\left\| {{\bf{w}}_0} \right\|^2} \le P \right\}, 
\end{aligned}
\end{equation}
where ${\bf{G}}\in \mathbb{C}^{M \times N_t}$, ${{\bf{h}}_{\rm{l}}} \in \mathbb{C}^{M \times 1}$, and ${{\bf{h}}_{\rm{e}}} \in \mathbb{C}^{M \times 1}$ denote the channels from the transmitter to the IRS and from the IRS to the legitimate receiver and eavesdropper, respectively. The transmit beamforming vector is denoted by ${\bf{w}}\in \mathbb{C}^{N_t \times 1}$ and $P \ge 0$ is the given transmit power. ${\sigma _{\rm{l}}^2}$ and ${\sigma _{\rm{e}}^2}$ are the variances of additive complex Gaussian noises at legitimate receiver and eavesdropper, respectively.
\par When the phase shift matrix ${\bf{\Phi }}$ is fixed, the optimal closed-form solution for beamformer ${\bf{w}}$ is given by \cite[Lemma 1]{Alex2019secure}. Given the beamforming vector, the original problem can be rewritten as
\begin{equation}\label{eq6}
\begin{aligned}
\max_{\bf{v}} \quad & f({\bf{v}})=\frac{{{{\bf{v}}^H}{{\bf{Y}}_{\rm{l}}}{\bf{v}}}}{{{{\bf{v}}^H}{{\bf{Y}}_{\rm{e}}}{\bf{v}}}}\\
\mbox{s.t.}\quad
&\left| {{{\bf{v}}_k}} \right| = 1, \quad k \in \{ 1,2, \cdots ,M\}, 
\end{aligned}
\end{equation}
where ${\bf{v}} = {[{e^{j{\theta _1}}},{e^{j\theta 2}},\cdots,{e^{j{\theta _M}}}]^H}$ and ${{\bf{Y}}_i} = \frac{1}{M}{{\bf{I}}_M} + \frac{1}{{\sigma _i^2}}{\rm{diag}}({\bf{h}}_i^H){\bf{Gw}}{{\bf{w}}^H}{{\bf{G}}^H}{\rm{diag}}({\bf{h}}_i^H)^H$, $i \in\{{\rm{l}}, {\rm{e}}\}$. According to \eqref{eq6}, the gradient of the objective function with respect to ${\bf{\Theta}} = {[{\theta _1},{\theta _2},\cdots,{\theta _M}]^T}$ is easily computable and given by
\begin{equation} \label{eq7}
\begin{aligned}
{\nabla _{\bf{\Theta}}}f = & 2{\rm{Re}} \left\{
\frac{{({{\bf{Y}}_{\rm{l}}}^*{{\bf{v}}^*}) \odot ( - j{\bf{v}})}}{{{{\bf{v}}^H}{{\bf{Y}}_{\rm{e}}}{\bf{v}}}} \right\}
\\
& + 2{\rm{Re}} \left\{ \frac{{({{\bf{v}}^H}{{\bf{Y}}_1}{\bf{v}}) \cdot ({{\bf{Y}}_{\rm{e}}}^*{{\bf{v}}^*}) \odot (j{\bf{v}})}}{{{{({{\bf{v}}^H}{{\bf{Y}}_{\rm{e}}}{\bf{v}})}^2}}} \right\}. 
\end{aligned}
\end{equation}
With \cite[Lemma 1]{Alex2019secure} representing Step 3 and \eqref{eq7} representing Step 4, the proposed AO-GD is well prepared.

\subsection{Weighted Sum-Rate Maximization for IRS-Aided Wireless Networks}
\par The second application investigates the IRS-aided multiple-input single-output (MISO) multiuser downlink communication systems \cite{guo2020weighted}. We aim at maximizing the weighted sum-rate (WSR) of the users by jointly optimizing the beamforming at the access point (AP) and phase shifts of IRS. Let ${\bf{h}}_{{\rm{d}},k} \in \mathbb{C}^{N_t \times 1}$, ${\bf{h}}_{{\rm{r}},k} \in \mathbb{C}^{M \times 1}$, and ${\bf{G}} \in \mathbb{C}^{M \times N_t}$ be channels from AP to user $k$, from IRS to user $k$, and from AP to IRS, respectively.
The received SINR of $k$-th user can be written as 
\begin{equation} \label{eq8}
{\gamma _k} = \frac{{{{\left| {\left( {{\bf{h}}_{{\rm{d}},k}^{H} + {{\bf{v}}^{H}}{{\bf{H}}_{{\rm{r}},k}}} \right){{\bf{w}}_k}} \right|}^2}}}{{\sum\limits_{i = 1,i \ne k}^K {{{\left| {\left( {{\bf{h}}_{{\rm{d}},k}^{H} + {{\bf{v}}^{H}}{{\bf{H}}_{{\rm{r}},k}}} \right){{\bf{w}}_i}} \right|}^2}}  + \sigma _0^2}},
\end{equation}
where ${\bf{v}} = {[{e^{j{\theta _1}}},{e^{j\theta 2}},\cdots,{e^{j{\theta _M}}}]^H}$ and ${{\bf{H}}_{{\rm{r}},k}} = {\mathop{\rm diag}\nolimits} \left( {{\bf{h}}_{{\rm{r}},k}^{\rm{H}}} \right){\bf{G}} \in \mathbb{C}^{M \times N_t}$. ${{\bf{w}}_k} \in \mathbb{C}^{N_t \times 1}$ is the transmit beamforming vector. $n_{k} \sim \mathcal{C} \mathcal{N}\left(0, \sigma_{0}^{2}\right)$ denotes the additive white Gaussian noise at the $k$-th receiver.

Let ${\bf{W}} = \left[ {{{\bf{w}}_1},{{\bf{w}}_2}, \cdots ,{{\bf{w}}_K}} \right] \in \mathbb{C}^{N_t \times K}$. The WSR maximization problem is formulated as
\begin{alignat}{2}
\max_{{\bf{W}},{\bf{v}}} \quad & {f}({\bf{W}},{\bf{\Phi }}) = \sum_{k=1}^{K} \omega_{k} \log \left(1+\gamma_{k}\right) & \label{eq9} \\
\mbox{s.t.}\quad
&\left| {{{\bf{v}}_i}} \right| = 1, \quad i \in \{ 1,2, \cdots ,M\}, & \tag{11a}\label{eq9a}\\
& {\bf{W}} \in {{\cal X}_1} = \left\{ \left. {\bf{F}} \right| \sum_{k=1}^{K}\left\|\mathbf{f}_{k}\right\|^{2} \leq P \right\}, & \tag{11b}\label{eq9b}
\end{alignat}
where the weight $\omega_{k}$ denotes the priority of user $k$ and $P$ is the downlink transmit power.

\par To solve Problem \eqref{eq9}, we apply the closed-form FP approach \cite{shen2018FP} to transform the sum-of-logarithms-of-ratio problem into the following equivalent problem
\begin{alignat}{2}
\max_{{\bf{p}}, {\bf{q}}, {\bf{W}},{\bf{v}}} \quad & {f_2}({\bf{p}}, {\bf{q}}, {\bf{W}},{\bf{v}}) & \notag \\
\mbox{s.t.}\quad
&{p_k} \ge 0,\quad \forall k = 1, \cdots ,K , & \notag \\
& \eqref{eq9a}, \eqref{eq9b}, & \notag
\end{alignat}
where ${\bf{p}}$ and ${\bf{q}}$ are auxiliary variables, and the new objective function is given in \cite{guo2020weighted}. One can refer to \cite[Eq. (11)-(12)]{guo2020weighted} to acquire the update rules of $p_k$ and $q_k$. When the auxiliary variables and reflection matrix of IRS are fixed, ${\bf{W}}$ is updated by solving $\mathop{{\rm{min}}}\limits_{{\bf{w}} \in {{\cal X}_1}} {f_3}({\bf{W}}) = {f_2}({{\bf{\bar p}}}, {{\bf{\bar q}}}, {\bf{W}},{{\bf{\bar v}}})$, where ${{\bf{\bar p}}}$, ${{\bf{\bar q}}}$, and ${{\bf{\bar v}}}$ denote the temporal optimization results in last block. We adopt a variant of proximal update rule to update ${{\bf{W}}}$ in closed-form solution. Detailed procedures can be found in \cite[Eq. (13)-(14)]{guo2020weighted} with extrapolation weight $\epsilon = 0$.
Given a fixed ${\bf{\Phi}}$, we iteratively update ${\bf{p}}$, ${\bf{q}}$ and ${\bf{W}}$ until stopping criterion $\epsilon_1$ triggers. These procedures form Step 3 in our general framework.

\par Given ${\bf{\bar p}}$, ${\bf{\bar q}}$, and ${\bf{\bar W}}$, ${\bf{v}}$ is optimized through
\begin{equation} \label{eq12}
\begin{aligned}
\min_{{\bf{v}}} \quad & {f_4}({\bf{v}}) =  {{\bf{v}}^H}{\bf{Rv}} - 2{\mathop{\rm Re}\nolimits} \left\{ {{{\bf{v}}^H}{\bf{e}}} \right\}\\
\mbox{s.t.}\quad
& \eqref{eq9a},
\end{aligned}
\end{equation}
where ${\bf{R}}$ and ${\bf{e}}$ are
\begin{alignat}{2}
{\bf{R}} & = \sum\limits_{k = 1}^K {{{\left| {{{\bar q}_k}} \right|}^2}} \sum\limits_{i = 1}^K {{{{\bf{\bar a}}}_{i,k}}} {\bf{\bar a}}_{i,k}^H, \notag \\
{\bf{e}} & = \sum\limits_{k = 1}^K {\left( {\sqrt {{\omega _k}\left( {1 + {{\bar p}_k}} \right)} \bar q_k^*{{{\bf{\bar a}}}_{k,k}} - {{\left| {{q_k}} \right|}^2}\sum\limits_{i = 1}^K {\bar b_{i,k}^*} {{{\bf{\bar a}}}_{i,k}}} \right)}, \notag
\end{alignat}
with ${{{\bf{\bar a}}}_{i,k}} = {{\bf{H}}_{{\rm{r}},k}}{{{\bf{\bar w}}}_i}$ and ${{\bar b}_{i,k}} = {\bf{h}}_{{\rm{d}},k}^H{{{\bf{\bar w}}}_i}$. Denote ${\bf{\Theta}} = {[{\theta _1},{\theta _2},\cdots,{\theta _M}]^T}$ as expressed in Section \uppercase\expandafter{\romannumeral2}. The gradient is given by
\begin{equation} \label{eq13}
{\nabla _{\bf{\Theta}}}f_4  = 2{\mathop{\rm Re}\nolimits} \left\{ {{\left( {{\bf{Rv}} - {\bf{e}}} \right)}^* \odot ( - j{\bf{v}})} \right\},
\end{equation}
which constitutes Step 4 in \textbf{Algorithm 1}. Overall, we first cyclically update ${\bf{p}}$, ${\bf{q}}$ and ${\bf{W}}$ until convergence, and then use the gradient information \eqref{eq13} to update ${\bf{\Phi}}$.

\section{Simulation Results}
\subsection{Secure Wireless Communications via IRS}
\begin{figure}[t] 
\centering
\includegraphics[height=5.7cm]{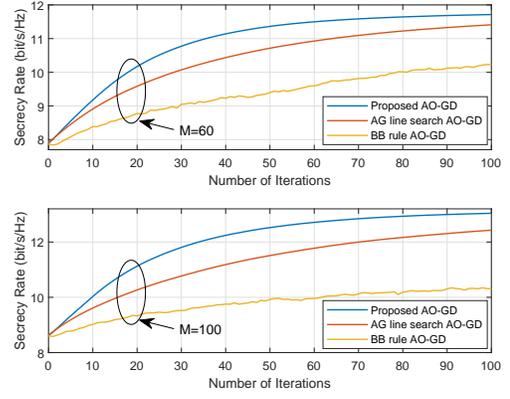} 
\caption{Convergence of different step size chosen schemes for different values of $M$ when $N_t=5$, $P=5$ dBm, $\alpha = 4$, ${r_{{\rm{TR}}}=250}$ m, and ${r_{{\rm{Rl}}}=r_{{\rm{Re}}}=160}$ m. (a) $M=60$. (b) $M=100$.} 
\label{figure1} 
\end{figure}
\par In this application, the distance between the transmitter and IRS is denoted by ${r_{{\rm{TR}}}}$, while ${r_{{\rm{Rl}}}}$ and ${r_{{\rm{Re}}}}$ stand for the distances from IRS to the legitimate receiver and the eavesdropper, respectively. Let $\alpha$ be the path loss exponent. All the other simulation settings are identical as those in \cite{Alex2019secure}. All algorithms start from a random initial point and the stopping criterion is that the increment of the normalized objective function value is less than $\xi=10^{-6}$. The simulation results are averaged over 1000 channel realizations. In \textbf{Algorithm 1}, we set the initial step size as $\gamma_0=0.001$, the step size decay factor as $\beta=0.5$, and $c=0.00005$. 

\par First, the convergence of the proposed algorithm is evaluated in Fig. \ref{figure1} compared with traditional step size chosen strategies, i.e., the AG line search \cite{numerical2006} and Barzilai-Borwein (BB) rule \cite{BB2005}. It is shown that our proposed AO-GD algorithm converges faster than traditional schemes. As the value of $M$ increases, i.e., from $M=60$ to $M=100$, the gap is more evident, justifying the efficiency of our framework for large-scale problems.

We then compare the performance of the proposed algorithm with other benchmarks in this problem. Specifically, the following 3 benchmarks are considered:
\begin{itemize}
\item \textbf{Element-wise BCD} \cite{Alex2019secure}: The phase shift matrix is updated by element-wise block coordinate descent. 
\item \textbf{AO-SDR} \cite{cui2019secure}: We leverage SDR and Gaussian randomization techniques to optimize phase shifts.
\item \textbf{AO-Manopt}: The stationary point of \eqref{eq6} is obtained via manifold optimization \cite{boumal2014manopt}.
\end{itemize}

\begin{figure}[t] 
\centering
\includegraphics[height=5.7cm]{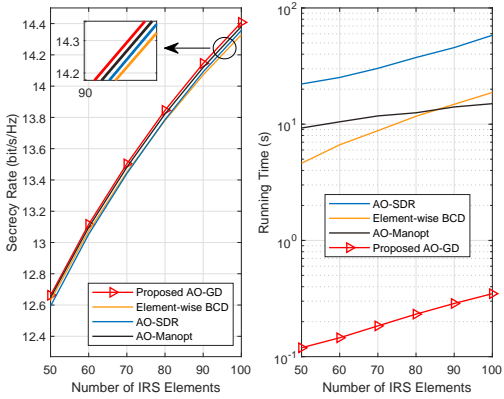} 
\caption{Average secrecy rate versus $M$ when $N_t=10$, $P=5$ dBm, $\alpha = 4$, ${r_{{\rm{TR}}}=200}$ m, ${r_{{\rm{Rl}}}=150}$ m, ${r_{{\rm{Re}}}=100}$ m.} 
\label{figure2}
\end{figure}
\par Fig. \ref{figure2} (left-hand side) shows the average secrecy rate of different algorithms versus the number of IRS elements. It is observed that the proposed AO-GD achieves the best performance in the whole regime. 
Besides, the average running time is plotted in Fig. \ref{figure2} (right-hand side). It is shown that all three benchmarks are time-consuming, which defers their use in large-scale IRS systems. The proposed framework requires the least running time, with more than 10 times speedup compared with the benchmarks. Overall, Fig. \ref{figure2} shows that the proposed framework is capable of obtaining high-performant solutions while enjoying high computational efficiency.

\subsection{Weighted Sum-Rate Maximization for IRS-Aided Wireless Networks}
\par In this application, we consider an AP equipped with 4 antennas, and 4 single-antenna users. 
All the other simulation parameters are identical as those in \cite{guo2020weighted}. The stopping criterion of the inner loop is $\xi_1=10^{-5}$ and the outer loop is $\xi_2=10^{-3}$, respectively. All the simulation results are averaged over 1000 channel realizations with random initialization.
\par In \textbf{Algorithm 1}, we set $\gamma_0=100$, $c=0.0001$, and step size decay factor $\beta=0.5$. We compare the performance of the proposed algorithm with the following 3 baselines:
\begin{itemize}
\item \textbf{Low-Complexity BCD} \cite{guo2020weighted}: \eqref{eq9} is decomposed into four disjoint blocks and the non-convex BCD method is applied to carry out the stationary solution to \eqref{eq9}.
\item \textbf{FP-SDR}: We leverage SDR and Gaussian randomization techniques to optimize phase shifts.
\item \textbf{FP-Manopt}: The stationary point of \eqref{eq12} is obtained via manifold optimization \cite{boumal2014manopt}.
\end{itemize}

\begin{figure}[t] 
\centering
\includegraphics[height=5.7cm]{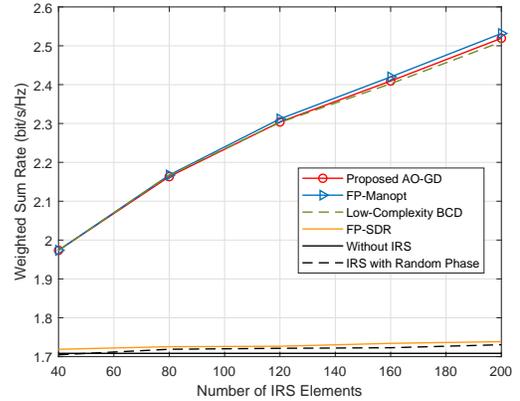} 
\caption{Average WSR versus $M$ when $P = 10$ dBm.} 
\label{figure3}
\end{figure}

\par Fig. \ref{figure3} demonstrates the WSR of different approaches with respect to the size $M$ of IRS when the transmit power is fixed as $P=10$ dBm. We observe that the SDR-based algorithm has only a small gain which is similar to the random phase scheme. This is because the solution obtained by SDR is almost full-rank in this case, leading to a poor performance after randomization. All the other schemes achieve comparable performance and remarkable gain as $M$ increases.
\par
\begin{figure}[t] 
\centering
\includegraphics[height=5.7cm]{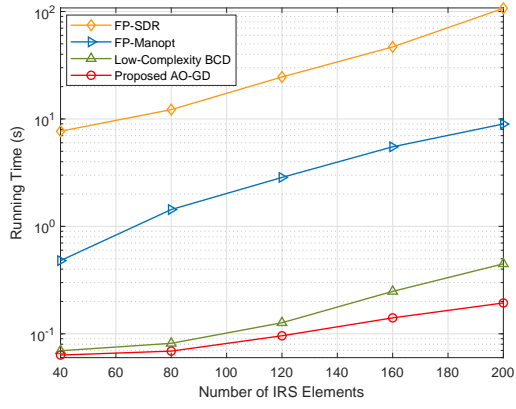} 
\caption{Average running time versus $M$ when $P = 10$ dBm.} 
\label{figure4} 
\end{figure}

 Fig. \ref{figure4} plots the average running time of different methods. It is shown that Manopt and SDR-based algorithms are time-consuming and their running time increases enormously as $M$ increases. It is observed that the proposed framework not only achieves significant speedup compared with FP-SDR and FP-Manopt, but also faster than the problem-specific algorithm, i.e., Low Complexity BCD. The speedup is more significant as the problem size increases, which demonstrate the scalability of the proposed framework.

\section{Conclusions}
This paper proposed a low-complexity algorithmic framework incorporating alternating optimization and gradient-based methods for large-scale IRS-assisted wireless systems. A general formulation and an algorithmic framework were firstly developed, followed by the convergence proof. Two examples were provided to demonstrate the efficiency and effectiveness of the proposed framework. Overall, this paper provided a simple yet effective approach for large-scale IRS-assisted systems, which can be easily implemented to different applications and serves as a good baseline to evaluate more sophisticated algorithms.

\appendix
\section{}\label{appA}
\begin{IEEEproof}
	We assume that $f(\cdot, \cdot)$ is differentiable with continuous gradient and feasible set $\cal{X}$ is compact. We know from Step 6 in \textbf{Algorithm 1} that $\{ f({{\bf{X}}^{(t)}})\} $ is a monotonically decreasing sequence which is bounded from below, and it thus converges. Since set $\cal X$ is bounded, $\{ {{\bf{X}}^{(t)}}\}$ has a convergent subsequence. Consider a limit point ${\bf{Z}}$ and a subsequence ${\{ {{\bf{X}}^{(t)}}\} _{t \in {T}}}$ converging to ${\bf{Z}}$, so we have
$\mathop {\lim }\limits_{T \ni t \to \infty } f({{\bf{X}}^{(t)}}) = \mathop {\lim }\limits_{T \ni t \to \infty } f({{\bf{X}}^{(t + 1)}}) = f({\bf{Z}})$.
Now we show that ${\bf{Z}}$ is a stationary point of Problem \eqref{general} based on \cite{InexactBCD}. On one hand, when ${\bf{\Phi}}^{(t)}$ is fixed, in every iteration, utilizing existing methods, ${\bf{Q}}^{(t)}$ is the stationary point of ${\rm{min}}_{{\bf{Q}} \in {{\cal X}_1}} \, {f}({{\bf{Q}} | {\bf{\Phi}}^{(t)}})$. Therefore, ${{\bf{Z}}_1}$ is the stationary point of ${\rm{\min}}_{{\bf{Q}} \in {{\cal X}_1}} \, {f}({{\bf{Q}} | {{\bf{Z}}_2}})$ and it satisfies the first-order optimality condition
\begin{equation}
{({\bf{Q}} - {{\bf{Z}}_1})^T}({\nabla _{\bf{Q}}}f({\bf{Z}})) \ge 0, \quad \forall {\bf{Q}} \in {{\cal X}_1}. \label{eq14}
\end{equation}
On the other hand, it follows from the step size chosen strategy that for all $t \in T$: $f({\bf{Q}}^{(t)},{\bf{\Phi}}^{(t+1)}) - f({\bf{Q}}^{(t-1)},{\bf{\Phi}}^{(t)}) \le - c{\gamma _0}{\beta ^{{m_t}}}{\nabla _{\bf{\Theta }}}f{({{\bf{X}}^{(t)}})^T}{\nabla _{\bf{\Theta }}}f({{\bf{X}}^{(t)}}) \le 0$,
where ${\gamma _0}{\beta ^{{m_t}}}$ denotes the step size at the $t$-th iteration, and hence
\begin{equation}
\mathop {\lim }\limits_{T \ni t \to \infty }{\beta ^{{m_t}}}{\nabla _{\bf{\Theta }}}f{({{\bf{X}}^{(t)}})^T}{\nabla _{\bf{\Theta }}}f({{\bf{X}}^{(t)}}) = 0. \label{eq15}
\end{equation}
From \eqref{eq15} we claim that 
\begin{equation}
\mathop {\lim }\limits_{T \ni t \to \infty } {\nabla _{\bf{\Theta }}}f({{\bf{X}}^{(t)}}) = {\bf{0}}. \label{eq16}
\end{equation}
To show this, we first assume the contrary: we assume $\mathop {\lim }\limits_{T \ni t \to \infty } {\nabla _{\bf{\Theta }}}f({{\bf{X}}^{(t)}}) = {\nabla _{\bf{\Theta }}}f({\bf{Z}}) \ne {\bf{0}}$.
Therefore, we have $\mathop {\lim }\limits_{T \ni t \to \infty } {\nabla _{\bf{\Theta }}}f{({{\bf{X}}^{(t)}})^T}{\nabla _{\bf{\Theta }}}f({{\bf{X}}^{(t)}}) > 0$,
and
\begin{equation}
\mathop {\lim }\limits_{T \ni t \to \infty } {\beta ^{{m_t}}} = 0.  \label{eq18}
\end{equation}
\eqref{eq18} indicates that there exists ${\tilde t}$ such that for $T \ni t \ge \tilde t$:
\begin{equation}
\begin{aligned}
& f({U}({{\bf{\Theta }}^{(t)}} - {\gamma_0}{\beta ^{{m_t} - 1}}{\nabla _{\bf{\Theta }}}f({{\bf{X}}^{(t)}})),{\bf{Q}}^{(t)})  \\
& > f({U}({{\bf{\Theta }}^{(t)}}),{\bf{Q}}^{(t - 1)}) - c{\gamma _0}{\beta ^{{m_t} - 1}}{\nabla _{\bf{\Theta }}}f{({{\bf{X}}^{(t)}})^T}{\nabla _{\bf{\Theta }}}f({{\bf{X}}^{(t)}}).
\end{aligned}
\notag
\end{equation}
Rearranging the terms we get
\begin{equation}\small \label{eq19}
\begin{aligned}
& \frac{{f({U}({{\bf{\Theta }}^{(t)}} - {\gamma_0}{\beta ^{{m_t} - 1}}{\nabla _{\bf{\Theta }}}f({{\bf{X}}^{(t)}})),{\bf{Q}}^{(t)}) - f({U}({{\bf{\Theta }}^{(t)}}),{\bf{Q}}^{(t - 1)})}}{{{\beta ^{{m_t} - 1}}}} \\ 
& > - c{\gamma _0}{\nabla _{\bf{\Theta }}}f{({{\bf{X}}^{(t)}})^T}{\nabla _{\bf{\Theta }}}f({{\bf{X}}^{(t)}}).
\end{aligned}
\end{equation}
Writing the Taylor expansion of $f({\bf{X}})$ at ${\bf{X}} = ({U}({{\bf{\Theta }}^{(t)}}),{\bf{Q}}^{(t - 1)})$ and letting $T \ni t \to \infty $, we obtain 
\[- {\nabla _{\bf{\Theta }}}f{({\bf{Z}})^T}{\nabla _{\bf{\Theta }}}f({\bf{Z}}) >  - c{\nabla _{\bf{\Theta }}}f{({\bf{Z}})^T}{\nabla _{\bf{\Theta }}}f({\bf{Z}}).\]
Since $0 < c < 1$, we can derive ${\nabla _{\bf{\Theta }}}f{({\bf{Z}})^T}{\nabla _{\bf{\Theta }}}f({\bf{Z}}) < 0$, which cannot be true. Therefore, \eqref{eq16} must hold, which implies $\angle ({{\bf{Z}}_2})$ is the stationary point of ${{\rm{min}}_{\bf{\Theta}}} \, f({U}({\bf{\Theta }}) | {{\bf{Z}}_1})$. Since problem \eqref{eq2} and problem \eqref{eq3} are equivalent, ${{\bf{Z}}_2}$ is the stationary point of ${{\rm{min}}_{\bf{\Phi}}} \, f({\bf{\Phi}} |  {{\bf{Z}}_1})$ with $\left| {{\bf{\Phi}}_{i,i}} \right| = 1, \forall i$, and it satisfies the first-order optimality condition
\begin{equation}
{({\bf{\Phi}} - {{\bf{Z}}_2})^T}({\nabla _{{\bf{\Phi}}}}f({\bf{Z}})) \ge 0, \quad \forall {\bf{\Phi}} \in {{\cal X}_2}. \label{eq20}
\end{equation}
Adding up \eqref{eq14} and \eqref{eq20}, we readily observe that ${\bf{Z}}$ satisfies first-order optimality condition, i.e., ${({\bf{X}} - {\bf{Z}})^T}(\nabla f({\bf{Z}})) \ge 0, \forall {\bf{X}} \in {\cal X}$.
The proof is thus completed.
\end{IEEEproof}

\bibliographystyle{IEEEtran}
\bibliography{IEEEabrv,technical_report_V6}

\end{document}